\documentclass{article}

\usepackage[main, final]{neurips_2026}

\usepackage[utf8]{inputenc} %
\usepackage[T1]{fontenc}    %
\usepackage{hyperref}       %
\usepackage{url}            %
\usepackage{booktabs}       %
\usepackage{amsfonts}       %
\usepackage{nicefrac}       %
\usepackage{microtype}      %
\usepackage{xcolor}         %
\usepackage{wrapfig}

\usepackage{amsthm}
\usepackage{amsmath}
\usepackage{amssymb}
\usepackage{cleveref}
\usepackage{graphicx}
\usepackage{subcaption}
\usepackage{stix}

\definecolor{tabblue}{RGB}{31,119,180}
\definecolor{taborange}{RGB}{255,127,14}

\title{{\it The Invisible Hand of Physics:} When Video Diffusion Models Know More Than They Show}

\author{
\textbf{Parsa Esmati}$^{*,1}$ \quad
\textbf{Somjit Nath}$^{*,2,3}$ \quad
\textbf{Katja Hofmann}$^{4}$ \\
\textbf{Derek Nowrouzezahrai}$^{2,3}$ \quad
\textbf{Samira Ebrahimi Kahou}$^{\dagger,3,5}$ \quad
\textbf{Majid Mirmehdi}$^{\dagger,1}$ \\[0.5em]
$^{1}$ University of Bristol \quad
$^{2}$ McGill University \quad
$^{3}$ Mila--Quebec AI Institute \\
$^{4}$ Microsoft Research \quad
$^{5}$ University of Calgary \\[0.3em]
{\small $^{*}$ Equal contribution \qquad $^{\dagger}$ Equal supervision}
}

\begin{document}

\maketitle

\begin{abstract}
Modern video diffusion models generate increasingly realistic and temporally coherent videos, motivating their use as candidate world simulators. Yet it remains unclear whether these models internally encode physical structure, or merely reproduce motion patterns seen during training. We study this question by probing video diffusion models along latent trajectories corresponding to real videos with known physical plausibility. To obtain such trajectories, we approximately invert the deterministic sampling process by integrating the learned velocity field backward from a clean video latent to noise, giving access to the model’s intermediate states and attention maps. Using these recovered trajectories, we show that physical plausibility is linearly decodable from diffusion transformer states across IntPhys and InfLevel, reaching around 81.27\% average accuracy and outperforming dedicated representation-learning baselines such as V-JEPA and VideoMAE. Surprisingly, this signal is absent from the VAE latent input and emerges inside the denoising transformer itself, despite the model not being trained with a self-supervised predictive objective. These findings suggest that physically meaningful representations can arise as a byproduct of generative denoising. Code is available \href{https://github.com/ParsaEsmati/videodiffusionphysics}{here}.
\end{abstract}

\section{Introduction}
The trajectory of video generation has been driven by a continual expansion of what these models are expected to capture. Early generative models~\citep{DBLP:journals/corr/VondrickPT16, DBLP:journals/corr/TulyakovLYK17} aimed to produce short, visually plausible clips, but the rise of large-scale diffusion-based generators~\citep{ho2022videodiffusionmodels} has shifted the goal toward modelling not only the visual aspects but also the dynamics of the visual world itself. Trained on internet-scale video, today's leading systems such as Sora~\citep{sora2024}, Veo~\citep{veo2024}, and Cosmos~\citep{nvidia2026worldsimulationvideofoundation} generate continuations of real scenes with striking temporal coherence, and are now being positioned as a path toward general-purpose simulators of the physical world for robotics, planning, and scientific discovery.

Despite these advances, it remains unclear whether such models have actually internalized the physical laws governing the scenes they generate. Visual realism alone does not require a model to represent acceleration as constant under gravity, momentum as conserved through a collision, or matter as impenetrable through contact; it require only that the model produces trajectories statistically similar to those it has seen during training. Recent evaluations~\citep{kang2025farvideogenerationworld, huberman2026semanticmomentstrainingfreemotionsimilarity} have shown that scaling video diffusion models fails to extrapolate basic mechanics outside the training distribution, with generations instead mimicking the nearest in-distribution example. Compounding this, large-scale video diffusion models operate in the latent space of a variational autoencoder (VAE) trained purely for reconstruction, which is not explicitly optimized to capture the semantic or physical structure that representation encoders are known to encode~\citep{bardes2024vjepa, garrido2025intuitivephysicsunderstandingemerges}. The model therefore has neither an explicit objective nor an implicit substrate that would push it to recover the laws governing the dynamics it is asked to generate.

This raises a fundamental question: {\it do modern video generation models encode physical knowledge internally, even when their output fails to capture it?}
Prior work has mainly approached this question through representation learning. Self-supervised encoders such as V-JEPA~\citep{bardes2024vjepa, assran2025vjepa2selfsupervisedvideo} learn latent spaces optimized for predicting future states and can distinguish physically plausible from implausible videos~\citep{garrido2025intuitivephysicsunderstandingemerges}. These results suggest that physical structure can emerge when models are trained with predictive objectives; however, it remains unclear whether a similar structure exists in diffusion models trained for generation rather than prediction.

A key obstacle is access: diffusion models do not expose the latent trajectories associated with real videos, making it difficult to probe how internal representations evolve. We address this by approximately inverting the generation process. Starting from a clean video latent, we integrate the learned velocity field backward to recover an approximate trajectory through the model’s intermediate states, providing access to its internal representations. Using this framework, we find that video diffusion models contain a clear, decodable signal of physical plausibility within their internal states, even when their generated outputs violate the same physical laws. This is surprising because these models are not trained with predictive or physics-aware objectives, and their inputs come from reconstruction-based VAEs that do not encode physical structure. Our results therefore show that physically meaningful representations can emerge as a byproduct of the denoising computation itself. Our analysis goes beyond probing by combining trajectory reconstruction with causal interventions, allowing us to study both where the physical information is readable and how it influences generation. 

Our contributions are threefold. First, we introduce a reverse-sampling approach to probing video diffusion models on real world videos. Second, we show that physical plausibility and quantitative physical variables are decodable from intermediate transformer blocks, despite having models trained without a predictive representation objective. Third, we characterize the structure of this signal across depth and causal interventions, showing that physical information emerges naturally and physical signal increases as trajectories more closely follow a model’s underlying continuous dynamics.

\section{Related Work}
We review prior works on physical understanding in video models, focusing on both generative approaches and self-supervised representation learning.

{\bf Video Generation and Physics.}
Diffusion-based models have driven rapid progress in video generation, enabling the synthesis of long, temporally coherent sequences with high visual fidelity~\citep{ho2022videodiffusionmodels, blattmann2023alignlatentshighresolutionvideo, sora2024, veo2024}. These models operate in latent spaces learned by pretrained autoencoders and generate videos by iteratively denoising noise samples. At scale, they exhibit emerging capabilities such as compositional reasoning and interaction understanding~\citep{wiedemer2025videomodelszeroshotlearners}, motivating their use as general-purpose world simulators. However, it remains unclear whether such behaviors reflect an internalization of physical laws or arise from statistical pattern matching over training data.

A growing body of work aims to improve the physical realism of generated videos. Some approaches introduce motion priors or enforce temporal consistency~\citep{ho2022imagenvideohighdefinition}, while others incorporate structured constraints or differentiable simulators~\citep{liu2025physflowunleashingpotentialmultimodal}. World-model-inspired methods instead attempt to learn dynamics directly in latent space, often using object-centric or factorized representations~\citep{DBLP:journals/corr/abs-1803-10122, DBLP:journals/corr/abs-1912-01603}. Despite these advances, recent evaluations show that current video models struggle to extrapolate even simple mechanics beyond their training distribution~\citep{huberman2026semanticmomentstrainingfreemotionsimilarity, kang2025farvideogenerationworld}. Importantly, this line of work focuses on improving outputs, rather than determining whether standard diffusion models already encode physical structure internally.

{\bf Representation Learning and Emergent Physics.}
A complementary line of work studies the emergence of physical understanding in learned representations. Self-supervised video encoders such as V-JEPA~\citep{bardes2024vjepa, assran2025vjepa2selfsupervisedvideo} learn predictive representations that encode semantic and dynamical structure. These representations have been shown to distinguish physically plausible from implausible videos with high accuracy~\citep{garrido2025intuitivephysicsunderstandingemerges}, suggesting that predictive objectives naturally encourage physical abstraction. Subsequent work has further localized these signals within model depth and features~\citep{joseph2026interpretingphysicsvideoworld}. In contrast, diffusion models rely on reconstruction-based latents and are not trained to predict future states, leaving it unclear whether comparable physical structure can emerge in their internal representations.

Concurrent work by \citet{tang2026seekingphysicsdiffusionnoise} reports similar physics-related cues in intermediate diffusion-transformer features of CogVideoX-2B, probed on clean latents with noise added directly. Under a deterministic sampler, however, each clean video is reached from a single noise sample along a single trajectory, so a random perturbation, especially at high noise levels, takes the probed state off that path. We instead follow the model's velocity field backward, so the probed states lie on the generative trajectory. We later demonstrate that the strength of the signal depends on how closely the probed states adhere to the true trajectory.

\section{Method}
\label{gen_inst}
\begin{figure}[t]
    \centering
    \begin{minipage}{0.6\linewidth}
        \centering
        \includegraphics[width=\linewidth]{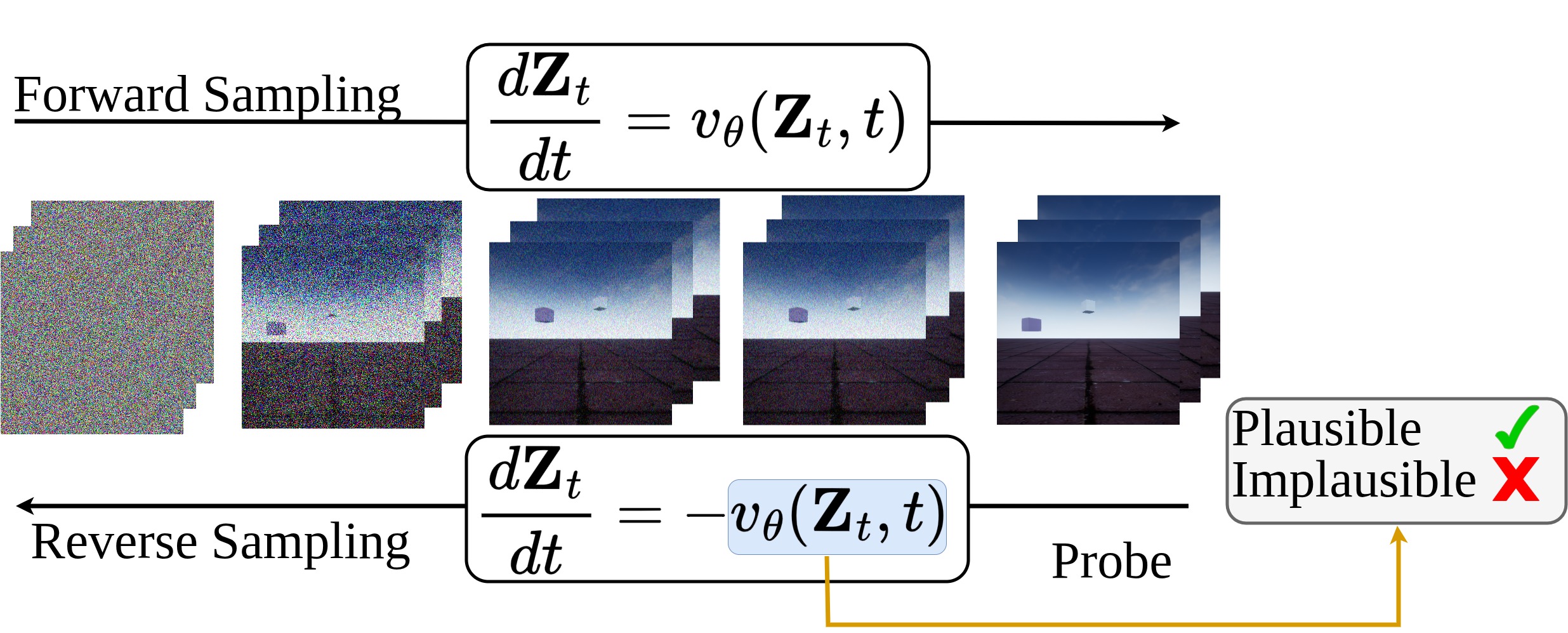}
    \end{minipage}%
    \hfill
    \begin{minipage}{0.38\linewidth}
        \caption{\textbf{Reverse sampling and probing.} Given a clean video latent $\mathbf{Z}_1$, we integrate the velocity field $v_\theta$ backwards to noise. Internal activations recorded at every block and timestep along the trajectory are probed to predict physical plausibility.}
        \label{fig:conceptual}
    \end{minipage}
\end{figure}

We present our approach for obtaining the internal features of a video diffusion model on any given real video. As illustrated in \Cref{fig:conceptual}, we recover an approximate latent trajectory by integrating the learned velocity field backward from a clean video latent to noise, and probe the internal activations along this trajectory. We first describe the flow-matching preliminaries and the forward sampling process from noise to clean data. We then present our reverse sampling procedure in Section~\ref{sec:reverse}, and quantify the approximation error that bounds the fidelity of the recovered internal representation in Appendix~\ref{app:reverse_error}. We then describe the block-level noise intervention and probe-surprise metric we use to identify which components are responsible for the physical signal in Section~\ref{sec:intervention}.

{\bf Preliminaries.} Flow-based generative models learn a time-dependent velocity field $v_\theta : \mathbb{R}^d \times [0, 1] \to \mathbb{R}^d$ that transports samples from a simple prior $p_0 = \mathcal{N}(\mathbf{0}, \mathbf{I})$ to the data distribution $p_1 \approx p_{\mathrm{data}}$. Given a noise sample $\mathbf{Z}_0 \sim p_0$, the model generates a clean sample $\mathbf{Z}_1$ by integrating the learned velocity field along the forward ordinary differential equation (ODE)
\begin{equation}
{d \mathbf{Z}_t}\big/{d t} = v_\theta(\mathbf{Z}_t, t), \qquad t \in [0, 1],
\label{eq:forward_ode}
\end{equation}
from $t = 0$ to $t = 1$. In practice, this integral is evaluated numerically with a fixed discretisation $0 = t_0 < t_1 < \cdots < t_N = 1$ and a one-step integrator such as Euler,
\begin{equation}
\mathbf{Z}_{t_{k+1}} = \mathbf{Z}_{t_k} + \Delta t_k \cdot v_\theta(\mathbf{Z}_{t_k}, t_k), \qquad \Delta t_k = t_{k+1} - t_k.
\label{eq:forward_euler}
\end{equation}
For a video diffusion model, $\mathbf{Z}_t$ is the latent encoding of a video under a pretrained autoencoder, and each evaluation of $v_\theta$ requires a full forward pass through a transformer backbone. This forward pass is the computation whose internal features we aim to examine.

\subsection{Reverse sampling}
\label{sec:reverse}
A core challenge in probing the internal representations of diffusion models is the lack of access to latent trajectories. The model provides no way to recover the internal state it would associate with a given real video, and so the trajectories we would want to examine are never produced. These are the trajectories tied to real videos with known physical plausibility, the only ones against which an internal signal can be tested. 

Our insight is that such trajectories can be recovered by running the model's own sampler in reverse. Concretely, for any video $X$ from a dataset of our choosing, we encode it with the variational autoencoder associated with the given video diffusion model to obtain the clean latent $\mathbf{Z}_1 = \mathcal{E}(X)$, and then integrate the forward ODE in reverse from $t = 1$ back to $t = 0$ to recover the noise sample $\mathbf{Z}_0$ together with the full trajectory $\{\mathbf{Z}_{t_k}\}_{k=0}^{N}$ traversed along the way.

The exact reverse of the Euler update in Eq.~\ref{eq:forward_euler} is the \emph{implicit} Euler step,
\begin{equation}
\mathbf{Z}_{t_k} = \mathbf{Z}_{t_{k+1}} - \Delta t_k \cdot v_\theta(\mathbf{Z}_{t_k}, t_k),
\label{eq:implicit_reverse}
\end{equation}
in which the unknown $\mathbf{Z}_{t_k}$ appears on both sides of the equation inside the nonlinear velocity field. Solving Eq.~\ref{eq:implicit_reverse} therefore requires an iterative solver at every step, multiplying the cost of reverse sampling by an order of magnitude over forward generation.

We find that an explicit approximation suffices: starting from $\mathbf{Z}_1$, we integrate the velocity field backwards to noise with an Euler or Heun scheme,
\begin{equation}
\mathbf{Z}_{t_k} = \mathbf{Z}_{t_{k+1}} - \Delta t_k \cdot v_\theta(\mathbf{Z}_{t_{k+1}}, t_{k+1}),
\label{eq:explicit_reverse}
\end{equation}
evaluating the velocity at the known endpoint instead of the unknown one. This requires a single network evaluation per step, matching the forward sampling cost. Resampling the recovered $\mathbf{Z}_0$ through the forward process recovers the original video up to minor artifacts, confirming that internal layers traverse video-preserving trajectories. We bound the approximation error in Appendix ~\ref{app:reverse_error}.

To quantify the physical information in these representations, we train linear probes~\citep{alain2018understandingintermediatelayersusing} on the recovered outputs of each of the transformer blocks to predict physical plausibility. Details of the probing protocol are provided in \Cref{sec:experiments}.

\subsection{Intervention}
\label{sec:intervention}

Having defined a way to extract internal representations via reverse sampling, we next ask which parts of the model are causally responsible for carrying that signal. Inspired by causal tracing and activation patching methods for localizing functional components in generative models~\citep{DBLP:journals/corr/abs-1906-04341,meng2023locatingeditingfactualassociations}, we perturb transformer blocks during generation and measure the resulting change in probe-assessed physical plausibility similar to~\citet{meng2023locatingeditingfactualassociations}. During generation, we hook into each transformer block in turn and corrupt its output activations, and then measure how much the resulting video's probe-assessed plausibility changes relative to an unmodified baseline generated from the same noise latent.

For each block, we replace its output hidden states $\mathbf{h}$ with
\begin{equation}
\tilde{\mathbf{h}} = \mathbf{h} + \alpha \cdot \boldsymbol{\sigma}(\mathbf{h}) \odot \boldsymbol{\epsilon}, \qquad \boldsymbol{\epsilon} \sim \mathcal{N}(\mathbf{0}, \mathbf{I}),
\label{eq:noise_intervention}
\end{equation}
where $\boldsymbol{\sigma}(\mathbf{h})$ is the per-token standard deviation over features and $\alpha$ is the intervention strength. Scaling noise to the local activation magnitude ensures $\alpha$ has consistent interpretation across blocks regardless of depth. The intervention is applied at every denoising step of the full generation trajectory.

Since diffusion models do not expose predictive future representations, we approximate surprise using the learned plausibility probes trained to classify physically plausible vs implausible videos from intermediate representations. These probes serve as a readout of the physical signal encoded in the model's internal blocks. To quantify how much a given block contributes to the physical signal, we adopt a \emph{probe-surprise} metric inspired by~\cite{garrido2025intuitivephysicsunderstandingemerges}, which interprets surprise as prediction error in representation space. In the absence of access to predictive latent trajectories, we use learned probes as a surrogate to estimate this error and thereby measure physical plausibility. 

For a generated video $V$, we re-invert it via the process in Section~\ref{sec:reverse}, capture hidden states at probe inversion steps $\mathcal{S}$, and score each with the corresponding linear probe. The probe surprise at step $s$ is
\begin{equation}
\psi_s(V) = \text{logit}_{\text{implausible}}(s, V) - \text{logit}_{\text{plausible}}(s, V),
\end{equation}
and we aggregate over steps as $\bar\psi(V) = \frac{1}{|\mathcal{S}|}\sum_{s\in\mathcal{S}} \psi_s(V)$. For each intervened video $V_b$, we record the surprise shift
\begin{equation}
\Delta_b = \bar\psi(V_b) - \bar\psi(V_{\text{base}}),
\end{equation}
where $V_{\text{base}}$ is the baseline video from the same noise latent without intervention. A positive $\Delta_b$ means corrupting block $b$ makes the video appear less physically plausible to the probe; a near-zero $\Delta_b$ means that block carries little physical signal.

\section{Experiments}
\label{sec:experiments}

Our experiments address five questions: (i) Is physical structure decodable from the internal states of video diffusion models, and how does this compare to representation-learning baselines such as V-JEPA (\Cref{sec:plausibility})? (ii) Where does this signal emerge across model depth and along the denoising trajectory (\Cref{sec:plausibility})? (iii) Which components of the model causally influence physical plausibility during generation (\Cref{sec:interventions})? (iv) Do these internal representations capture underlying physical variables beyond binary plausibility (\Cref{sec:box2d})? %
(v) Does preserving the reconstructed endpoint suffice, or does physical decodability depend on trajectory fidelity (\Cref{sec:steps})?

\subsection{Physical Understanding Emerges Internally}
\label{sec:plausibility}
\begin{figure}[t]
    \centering
    \includegraphics[width=\linewidth]{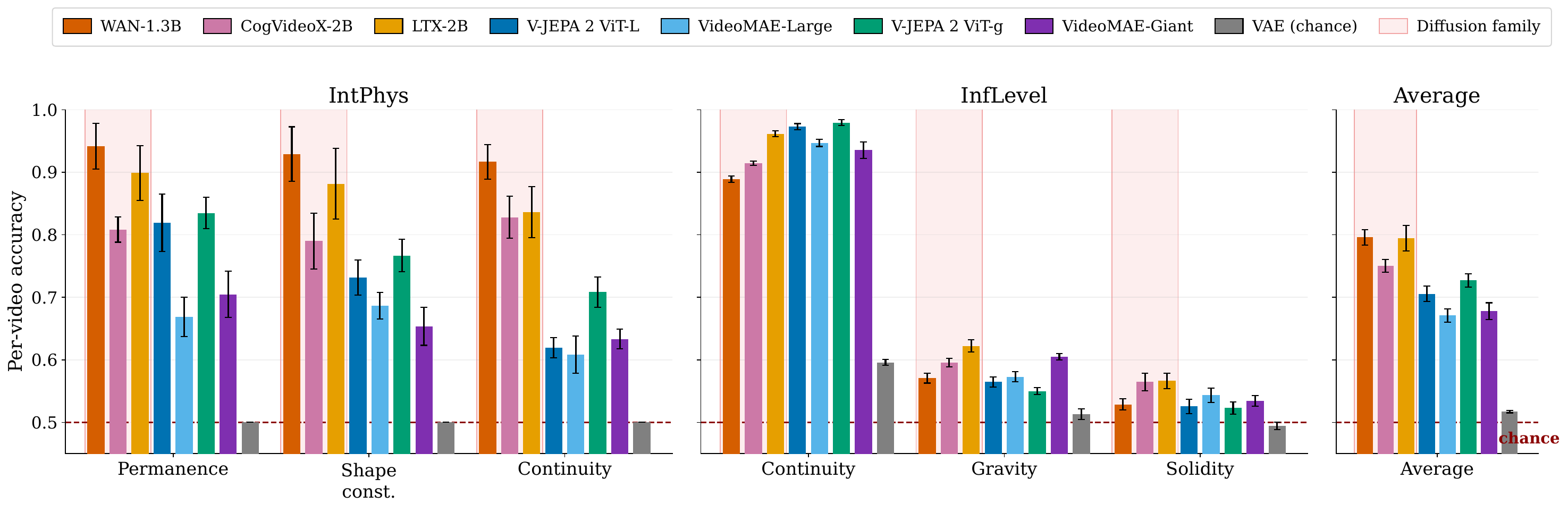}
    \caption{\textbf{Physical plausibility is decodable from the internal states of video diffusion models.} Probe accuracy on \textbf{(Left)} IntPhys and \textbf{(Middle)} InfLevel for WAN, LTX, and CogVideoX, compared against V-JEPA 2 (L,g) and VideoMAE (L,g). \textbf{(Right)} Diffusion video models outperform representation encoder on average. Error bars demonstrate standard error of the mean across 5 seeds.}
    \label{fig:main_plausibility}
\end{figure}

{\bf Benchmarks and baselines.} To enable a direct comparison with existing evidence on physical understanding in video models, we first evaluate on two benchmarks widely used in this literature: IntPhys~\citep{DBLP:journals/corr/abs-1803-07616} and InfLevel~\citep{WeihsEtAl2022InfLevel}. Both provide pairs of physically plausible and implausible videos and have been used to evaluate self-supervised representation encoders through the violation-of-expectation paradigm, which allows us to place our results alongside V-JEPA 2~\citep{bardes2024vjepa,assran2025v} and VideoMAEv2~\citep{wang2023videomaev2scalingvideo}. We additionally report results on the more recent IntPhys~2 benchmark~\citep{bordes2506intphys}, which applies the same paradigm to more complex settings (Appendix~\ref{app:intphys2}). Finally, we adopt the dataset from~\citet{kang2025farvideogenerationworld} to evaluate whether the internal representations carry information beyond a binary plausibility. This dataset is generated by a deterministic 2D physics simulator with known, controllable parameters including initial velocity, mass, size, and trajectory type, which allows us to test whether the model's internal states encode the underlying physical parameters rather than simple plausibility output. Datasets are further discussed in Appendix~\ref{app:datasets}.
 
{\bf Backbones.} We study three large-scale video diffusion models that operate in the latent space of normal VAEs: WAN~\citep{wan2025}, LTX~\citep{hacohen2024ltx}, and CogVideoX~\citep{hong2022cogvideo,yang2024cogvideox}. This covers three distinct architectural families within the current generation of latent video diffusion models, and allows us to assess whether our findings are specific to a particular backbone or reflect a more general property of how physical information is organized in diffusion-based video generators. Beyond the primary results presented in the next section for these three models, we also demonstrate the latent space structure in which they operate which does not demonstrate any physical structure (see Appendix~\ref{app:tsne}). 

{\bf Probing protocol.}
For each video, we recover the latent trajectory using the reverse sampling procedure of \Cref{sec:reverse}. At each inversion step and transformer block, we extract the hidden representations and train a linear probe to predict physical plausibility on a held-out test split. We use a 60/40 train/test split across all models and report held-out accuracy averaged over categories. Probes are trained independently for each block and inversion step, allowing us to map where in depth and along the trajectory physical information becomes linearly decodable.

{\bf Main comparisons.}
We first test whether physical plausibility is decodable from the internal states of a video diffusion model. For each video, we apply the reverse sampling procedure of \Cref{sec:reverse} with $K = 100$ integration steps, extract the internal activations at every transformer block along the trajectory, and train the probe to classify plausible from implausible clips on IntPhys and InfLevel. \Cref{fig:main_plausibility} reports mean probe accuracy across blocks at the midpoint of the reverse trajectory ($t = 0.5$), compared with mean probe accuracy of V-JEPA 2 (ViT-L and ViT-g)~\citep{bardes2024vjepa,assran2025v} and VideoMAE (Large and Giant)~\citep{wang2023videomaev2scalingvideo}. Across both benchmarks and all three video diffusion models we evaluate, namely WAN-1.3B, CogVideoX-2B, and LTX-2B, probe accuracy on internal states reaches and frequently exceeds that of dedicated representation encoders. Averaged across all categories of both benchmarks (\Cref{fig:main_plausibility} Right), every diffusion model outperforms all four encoders, with WAN-1.3B reaching $81.27\%$ against $71.36\%$ for V-JEPA 2 ViT-L.

\begin{wrapfigure}{r}{0.45\linewidth}
    \vspace{-10pt}
    \centering
    \includegraphics[width=\linewidth]{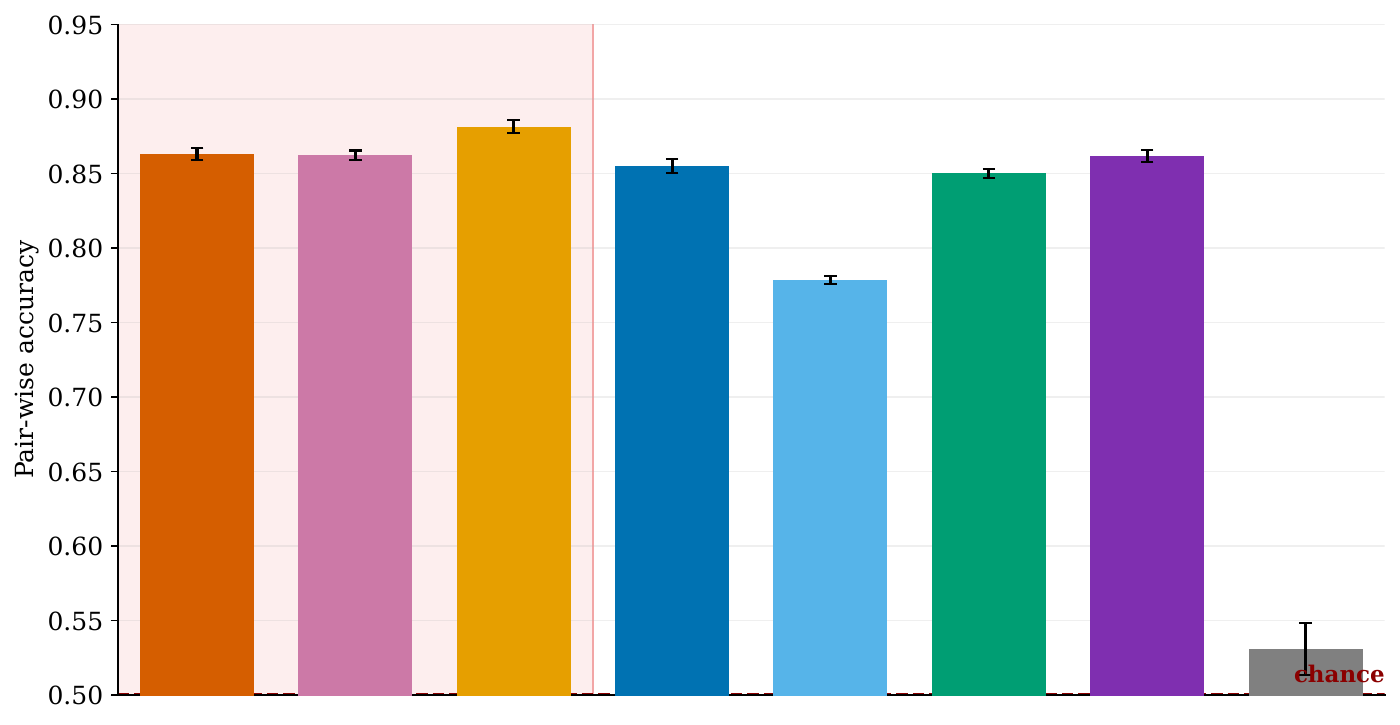}
    \vspace{-15pt}
    \caption{\textbf{Pairwise accuracy on IntPhys and InfLevel.} Following the protocol of~\citet{garrido2025intuitivephysicsunderstandingemerges}, the probe is shown a plausible/implausible pair and predicts which is implausible; colours as in \Cref{fig:main_plausibility}. The diffusion models match or exceed all four encoders (V-JEPA~2 ViT-L and ViT-g, VideoMAE-Large and VideoMAE-Giant).}

    \label{fig:pairwise_plausibility}
    \vspace{-10pt}
\end{wrapfigure}

Note that the accuracies reported in \Cref{fig:main_plausibility} are per-video, where the probe must classify each clip in isolation. As noted by~\citep{DBLP:journals/corr/abs-1803-07616, garrido2025intuitivephysicsunderstandingemerges}, this is a more difficult task than a simple pairwise evaluation where the model is shown a plausible/implausible pair sharing the same context and asked only which of the two is the impossible one. For completeness and a result comparable to the previous work, we also present the pairwise evaluation in  \Cref{fig:pairwise_plausibility}.

This result is non-trivial since the latent space these diffusion models operate in carries no physical signal on its own: probing the VAE latents directly, before any flow computation, yields chance accuracy (\~50\%) across all categories. This holds for all three VAEs we evaluate, so we report their averaged accuracy as a single value in \Cref{fig:main_plausibility}. We also show in Appendix~\ref{app:tsne} that the VAE latents of plausible and implausible videos are visually indistinguishable under t-SNE. The physical signal we recover is therefore not inherited from a structured input representation, as it is for V-JEPA whose embedding space is shaped by a self-supervised predictive objective. It is constructed by the diffusion model itself, in the course of transporting the video through its learned flow. This makes the result qualitatively different from prior findings in V-JEPA-style models, since the representation is not the training target but a byproduct of generative denoising.

\begin{figure}[ht]
    \centering
    \includegraphics[width=\linewidth]{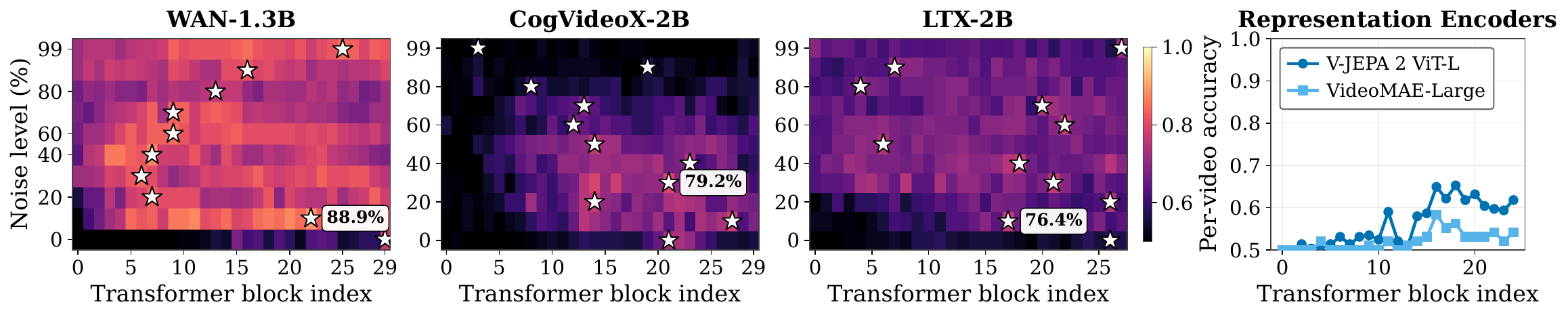}
    \caption{\textbf{Physical plausibility emerges across the entire reverse trajectory and at intermediate depth.} For each diffusion model (WAN-1.3B, CogVideoX-2B, LTX-2B), we report per-video probe accuracy at every transformer block (x-axis) and every noise level along the reverse trajectory (y-axis); $\bigwhitestar$ mark the best-performing block at each noise level. Right: per-block accuracy of V-JEPA 2 ViT-L and VideoMAE-Large under the same probing protocol, plotted on the same y-axis range.}
    \label{fig:emergence_zone}
\end{figure}

{\bf Emergence zone.}
One might assume that physical representations emerge only once the model is close to clean data, and that our choice of reporting at $t = 0.5$ in \Cref{sec:plausibility} obscures a strong dependence on the noise level. We test this directly by probing every transformer block at every point along the reverse trajectory, and mapping where, in depth and time, the physical signal is concentrated. \Cref{fig:emergence_zone} shows per-video probe accuracy across this full grid for all three diffusion models, alongside per-block accuracy of V-JEPA 2 ViT-L and VideoMAE-Large under the same protocol.

The physical signal is sustained across most of the reverse trajectory. For every diffusion model and every noise level except at 0.0\%, the best-performing block achieves an accuracy above $76\%$, and the best performing model WAN achieves $88.90\%$ in its richest block. Notably, these physical signals are much weaker in earlier blocks and near the clean data. Similarly, throughout noise levels, the first and second block commonly achieve lowest accuracies in the given noise level. Beyond this trivial floor, physical decodability is not localised to a narrow band of noise levels: the model maintains it from near-noise all the way to near-data.

The signal is, however, localised in depth. Across all three diffusion models the best blocks cluster nearly in the intermediate-to-late depth of the network, with the best block across noise levels at block $22$ for WAN-1.3B, block $21$ for CogVideoX-2B, and block $17$ for LTX-2B. This is consistent with the emergence zone reported by~\citet{joseph2026interpretingphysicsvideoworld} for V-JEPA, which we also reproduce in the right panel of \Cref{fig:emergence_zone}: physical information emerges at intermediate depth across architectures with otherwise different training objectives, suggesting that intermediate-depth emergence is a general property of how video models organise physical computation rather than an artefact of any particular training recipe. Thus, physical information is not confined to the clean-video endpoint; it is organized throughout the denoising trajectory, with strongest linear decodability at intermediate depth.

\subsection{Causal Localization via Block Interventions}
\label{sec:interventions}
\begin{figure}[t]
    \centering
    \includegraphics[width=\linewidth]{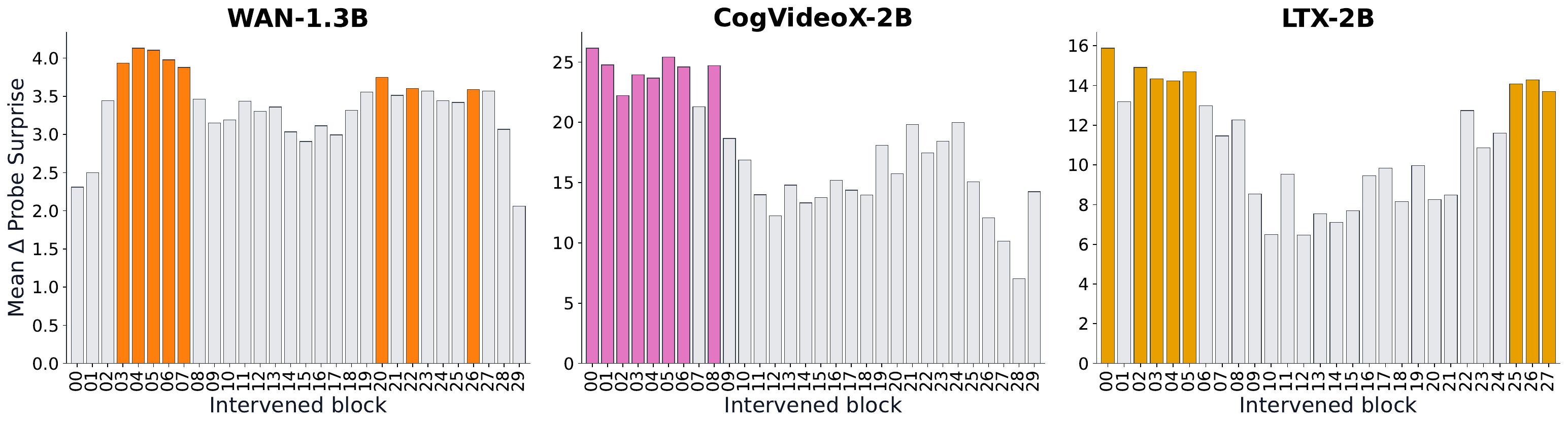}
    \caption{Mean probe-surprise shift $\Delta_b$ per transformer block on IntPhys, evaluated using the step 50 (noise level 50\%) probe. Colored bars highlight the top 8 blocks with highest surprise per model, while gray bars denote the remaining blocks. Larger values indicate that perturbing a given block leads to a greater degradation in the surprise probe metric.}
    \label{fig:intervention_bar}
\end{figure}

To identify which transformer blocks are causally responsible for the physical signal, we apply the noise intervention of Section~\ref{sec:intervention} independently to each of the 30 blocks of WAN across 180 plausible scenes, and measure the resulting surprise shift $\Delta_b$ averaged over scenes and probe inversion steps. Importantly, the blocks that are easiest to decode from need not be identical to the blocks whose corruption most affects generation. The former measures where physical information is most linearly readable; the latter measures which components causally influence the generated trajectory.

\Cref{fig:intervention_bar} shows that causal sensitivity is distributed across depth, with a structured and model-dependent pattern rather than a single localized region. For WAN-1.3B, early layers (blocks 0--8) exhibit the strongest sensitivity, with peak shifts around blocks 3--5, but mid and later layers retain non-negligible influence, leading to a relatively broad distribution of causal effect. CogVideoX-2B displays a sharper early-layer dominance, where the first few blocks produce the largest surprise shifts, followed by a drop in intermediate layers and a partial recovery at later depth. In contrast, LTX-2B exhibits a distinctly bimodal pattern: both early and late layers show high sensitivity, while intermediate layers contribute comparatively less.

These results indicate that causal influence over physical plausibility is not confined to a narrow subset of layers, but instead distributed across multiple stages of computation. Rather than being localized to specific layers, physical information appears to be injected early, transformed across depth, and in some architectures further refined at later stages. Comparing with the probing results of \Cref{sec:plausibility}, we observe that the layers where physical information is most linearly decodable do not coincide with those that are most causally sensitive. Probing peaks at intermediate depth, whereas interventions show that perturbing earlier (and in some cases later) layers has the largest effect on the generated trajectory. This suggests that physical structure is established early in the computation and propagated through the network, becoming most linearly accessible only at intermediate layers.

\subsection{Compressed DiT States Yield Stronger Probing Accuracy}
\label{sec:compression}
\begin{figure}[t]
    \centering
    \includegraphics[width=\linewidth]{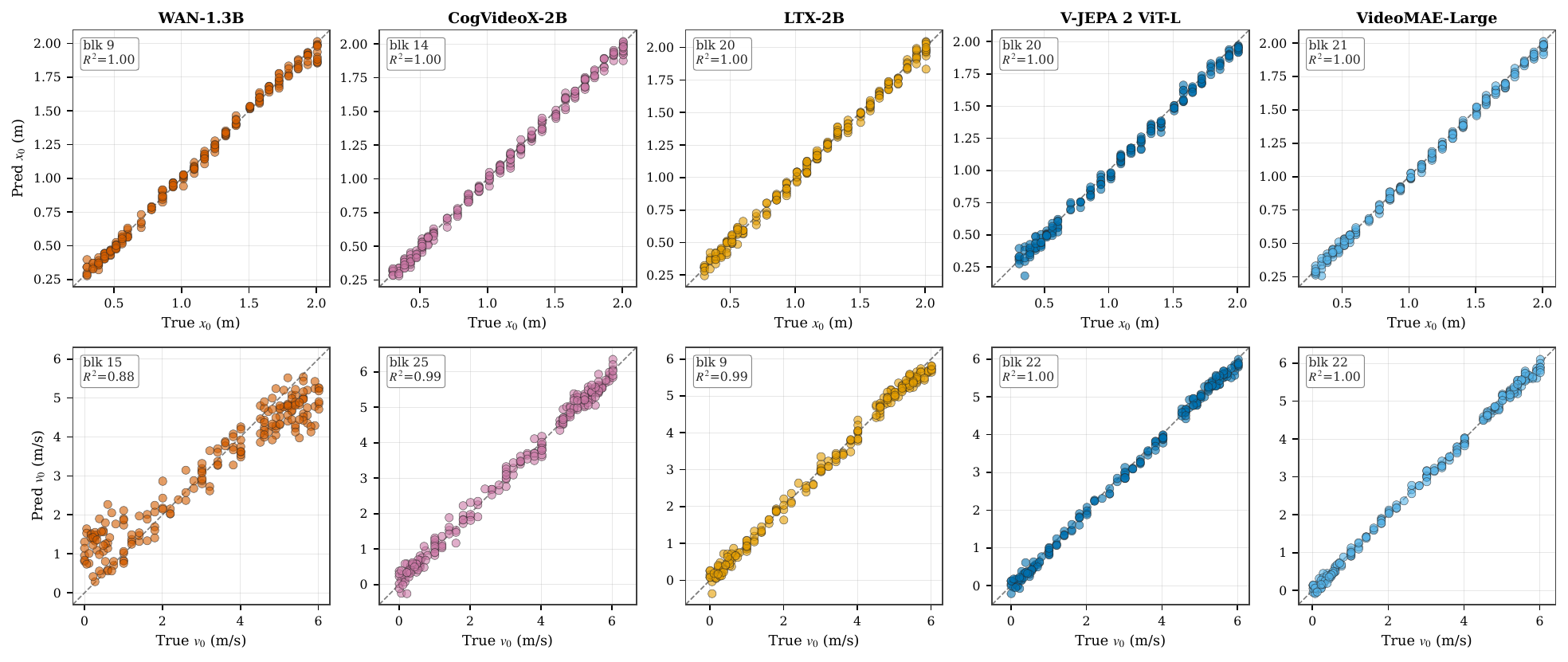}
    \caption{\textbf{Internal diffusion model states encode quantitative scene parameters.} For each model we train a linear regressor on all the blocks to predict the initial position $x_0$ (top) and initial velocity $v_0$ (bottom) of a parabolic ball trajectory~\citep{kang2025farvideogenerationworld} and plot the best-performing one. All three models match V-JEPA 2 and VideoMAE-Large on $x_0$, with $R^2 = 1.00$. For $v_0$, WAN-1.3B reaches $R^2 = 0.88$ while wider DiTs reach $R^2 \geq 0.99$, consistent with capacity arguments in \Cref{sec:compression}.}
    \label{fig:beyond_plausibility}
\end{figure}
\begin{wrapfigure}{r}{0.38\linewidth}
    \vspace{-12pt}
    \centering
    \includegraphics[width=\linewidth]{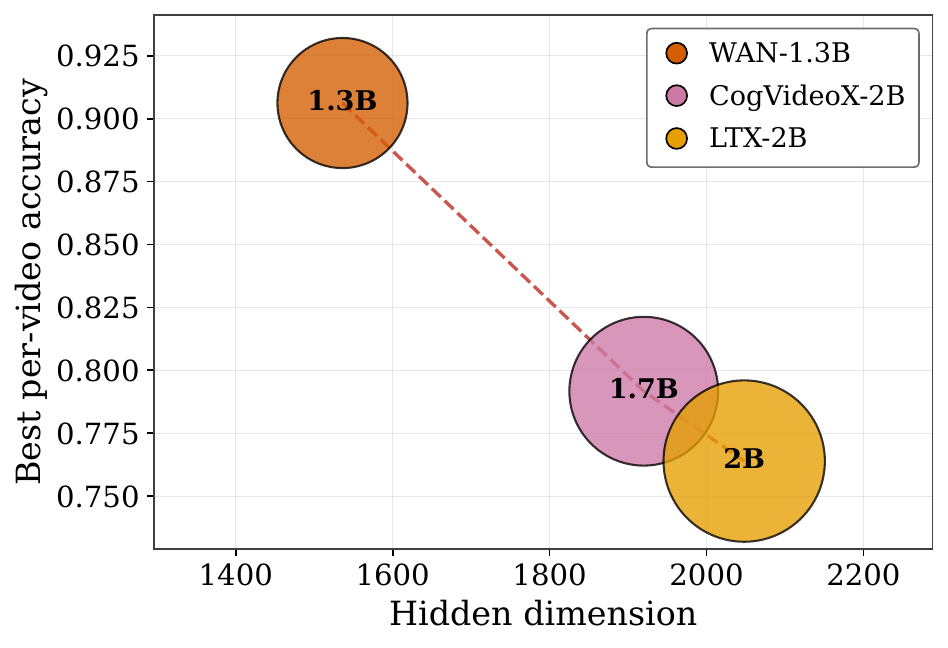}
    \vspace{-15pt}
    \caption{\textbf{Best per-video accuracy decreases with DiT dimension.} Across three diffusion models, probe accuracy at the best block is inversely related to the hidden dimension of the DiT, and is not explained by parameter count.}
    \label{fig:compression}
    \vspace{-15pt}
\end{wrapfigure}

A natural question is whether the gap between WAN-1.3B and the larger CogVideoX-2B and LTX-2B reflects a specific architectural choice or training recipe. Although this is a difficult aspect to isolate, we observe a linear relationship between the physical understanding and DiT dimensions. \Cref{fig:compression} plots the best per-video accuracy against the internal dimensions of the DiT. Accuracy decreases monotonically as the dimension grows, and the ranking does not track parameter count: WAN-1.3B is both the smallest model and the most compressed, and yields the highest probe accuracy.

One possible explanation is that this is a consequence of the limited representational capacity each DiT has available for transporting noise to clean video. A narrower model cannot encode every aspect of the scene equally well, and is forced to prioritise the spatiotemporal semantic structure required for coherent denoising over the high-frequency texture detail that determines visual sharpness. A wider model has the capacity to do both, and its internal states correspondingly allocate part of their dimensionality to texture-level content that is irrelevant to physical plausibility. The physical signal in a wider DiT is therefore not absent but diluted, spread across dimensions that simultaneously encode information the probe does not need. We treat this trend as suggestive rather than conclusive, since it is measured across three architectures that differ in more than hidden dimension.

\subsection{Beyond Binary Plausibility}
\label{sec:box2d}

Benchmarks (\Cref{sec:plausibility}) test whether the model can distinguish a physically plausible video from an implausible one, which is a binary judgement. We next ask whether the internal signal reflects genuine physical variables rather than only a binary plausibility boundary. We use the parabolic-motion subset of~\citet{kang2025farvideogenerationworld}, in which a ball is launched from a known initial position $x_0$ and initial velocity $v_0$ and evolves under deterministic 2D physics. For each video we extract internal activations at every block at $t = 0.5$ and train a linear regressor to predict $x_0$ and $v_0$ from a single block.

\Cref{fig:beyond_plausibility} shows that all three diffusion models recover both quantities with near-perfect linear fit. Every model achieves $R^2 \geq 0.99$ on $x_0$, and the diffusion models reach $R^2$ values between $0.88$ and $0.99$ on $v_0$, matching V-JEPA 2 and VideoMAE-Large. The internal states therefore carry not only the fact that the trajectory is consistent with gravity but the specific initial conditions that generated it. These parameters are not explicitly supervised during diffusion training.

The one informative deviation is on $v_0$, where WAN-1.3B reaches $R^2 = 0.88$ against $0.99$ to $1.00$ for the larger models. This pattern is consistent with the capacity argument of \Cref{sec:compression}. A narrow DiT is forced to allocate its limited representational budget to the coarse spatiotemporal semantics that determine whether a trajectory is physically plausible. These semantics are sufficient to support binary plausibility detection and to recover a positional quantity like $x_0$ that varies on a slow scale across the video. Recovering an instantaneous quantity like $v_0$ requires resolving finer-grained temporal detail, and a wider DiT has the capacity left over after encoding the coarse semantics to preserve enough of this detail to support precise regression. The same property that gives WAN the cleanest binary physical signal also limits how finely it can read off the parameters underlying that signal.

\subsection{Physics lives in the flow}
\label{sec:steps}

\begin{wrapfigure}{r}{0.4\linewidth}
    \vspace{-12pt}
    \centering
    \includegraphics[width=\linewidth]{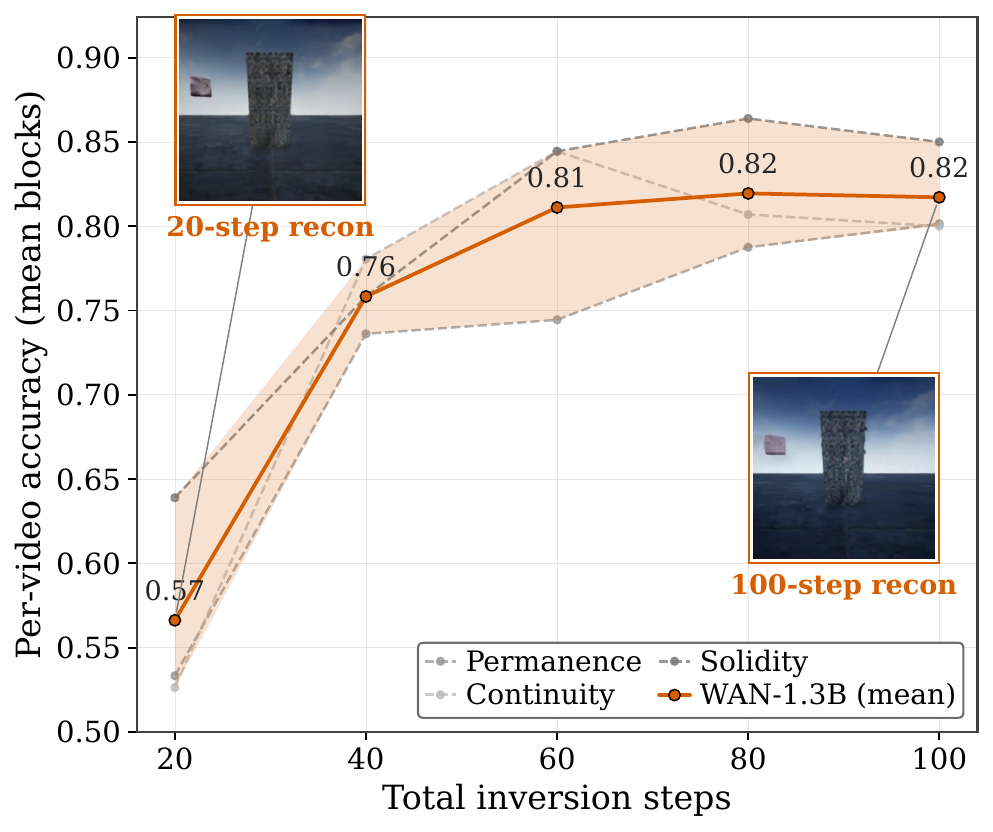}
    \vspace{-15pt}
    \caption{\textbf{Reconstruction can succeed where physical decoding fails.} Reconstructions from reverse sampling with $20$ and $100$ steps remain visually faithful, however per-video probe accuracy as a function of inversion steps collapses from $0.82$ at $100$ steps to $0.57$ at $20$.}
    \label{fig:steps}
    \vspace{-12pt}
\end{wrapfigure}

The reverse sampling procedure of \Cref{sec:reverse} approximates a continuous ODE with $K$ discrete steps.
We test whether physical signals are preserved as long as the reconstructed video remains visually faithful.
\Cref{fig:steps} reports both quantities as a function of $K$ for WAN-1.3B. The two come apart sharply. At $K = 20$ the recovered noise reconstructs the original video with only mild quality loss, yet probe accuracy collapses from $0.82$ to $0.57$, close to chance. At $K = 40$ accuracy already recovers most of the way to its asymptote, and the curve plateaus thereafter. The discretisation needed to generate a physically plausible-looking video is therefore strictly coarser than the discretisation needed to compute physical information.

We interpret this finding as evidence that the physical signal is a property of the \emph{exact ODE trajectory} the model would integrate in the continuum limit, not of the endpoints of that trajectory. Training the model to transport every noise sample to every clean video shapes a specific path through the latent space, and the internal states encode physical information at every point along that path. A coarse discretisation does not follow this path. It takes large steps that land in the same neighbourhood at every checkpoint, so the endpoints still produce a faithful video, but the intermediate states no longer correspond to the points the model would have visited under exact integration. The trajectory is what carries the physics, and a coarse trajectory is a different trajectory.

\begin{wrapfigure}{r}{0.38\linewidth}
    \centering
    \includegraphics[width=\linewidth]{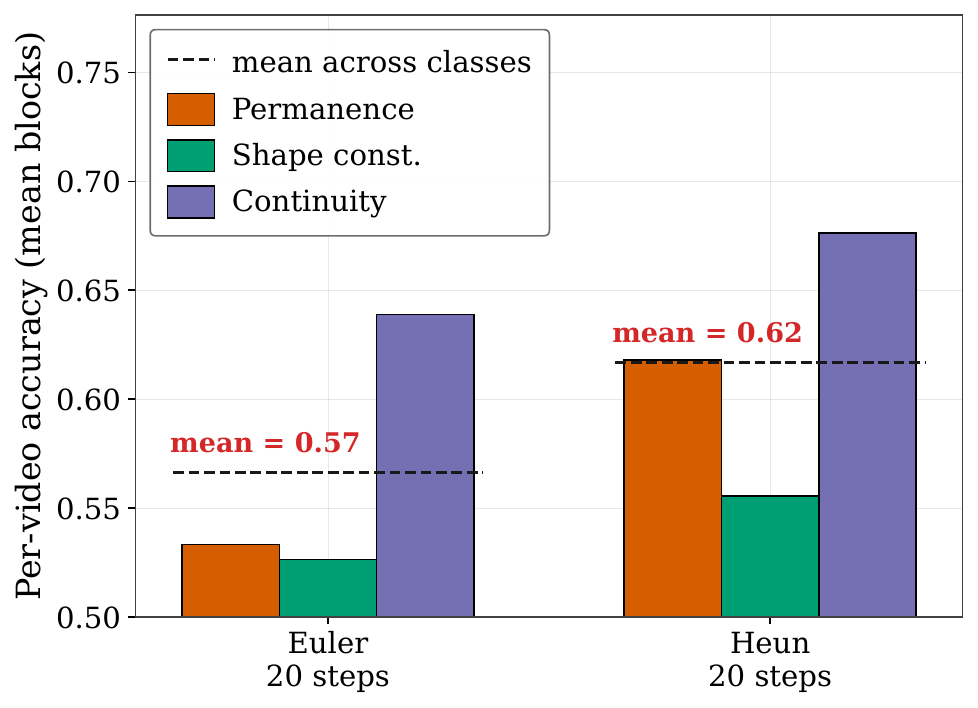}
    \vspace{-15pt}
    \caption{\textbf{Accurate discretization improves decodability.} Probe accuracy with Heun solver exceeds Euler.}
    \label{fig:integrator}
    \vspace{-40pt}
\end{wrapfigure}

To test this further we repeat the 20-step reverse sampling with a Heun solver and compare it directly to Euler. \Cref{fig:integrator} shows that the more accurate integrator partially recovers the physical signal, supporting the view that decodability tracks the fidelity of the discretised trajectory.

\section{Conclusion and Future Work}

We studied whether video diffusion models internally encode physical structure despite being trained purely for generation. By approximately inverting the sampling process, we recovered latent trajectories for real videos and probed the internal states of multiple diffusion models. We find that physical plausibility is linearly decodable from transformer states, even though it is absent from the VAE latent input. This signal persists across the reverse trajectory, is most accessible at intermediate depth, and encodes quantitative physical parameters. These results suggest that physically meaningful representations can emerge as a byproduct of generative denoising. Future work could leverage these signals for physics-aware guidance and further explore their potential role as latent spaces for world modeling.

\paragraph{Limitations} Our analysis relies on approximate reverse sampling, which may introduce trajectory errors. While reconstruction consistency supports its validity, fine-grained conclusions may depend on discretization. Linear probes demonstrate decodability but do not imply that the model explicitly represents physical laws, and our probe-based metric may inherit bias from the learned classifier. More broadly, our results are correlational: although interventions provide partial causal evidence, they do not fully establish that the model internally encodes physical laws in a mechanistic or human-interpretable form. 

\begin{ack}
This work was supported by the UK Engineering and Physical Sciences Research Council (EPSRC) under grant EP/W524414/1. This work was also supported by the Natural Sciences and Engineering Research Council of Canada (NSERC). The authors acknowledge the use of resources provided by the Isambard-AI National AI Research Resource (AIRR). Isambard-AI is operated by the University of Bristol and is funded by the UK Government’s Department for Science, Innovation and Technology (DSIT) via UK Research and Innovation; and the Science and Technology Facilities Council [ST/AIRR/I-A-I/1023].

\end{ack}

\bibliographystyle{plainnat}
\bibliography{neurips_2026}

\appendix
\section{Error Analysis of Explicit Reverse Sampling}
\label{app:reverse_error}

We derive the local and global error of the explicit reverse sampling scheme (Eq.~\ref{eq:explicit_reverse}) relative to the exact implicit inverse (Eq.~\ref{eq:implicit_reverse}). We reuse the notation of Section~\ref{sec:reverse}: $v_\theta$ is the learned velocity field, $\{t_k\}_{k=0}^{N}$ is a uniform time grid with step size $h = t_{k+1} - t_k$, and $\mathbf{Z}_{t_k}^{\mathrm{imp}}$, $\mathbf{Z}_{t_k}^{\mathrm{exp}}$ denote the implicit and explicit reverse trajectories, both initialised at $\mathbf{Z}_1$. We assume $v_\theta$ is $C^1$ in both arguments and $L$-Lipschitz in its first argument.

\paragraph{Local error.}
Subtracting Eq.~\ref{eq:implicit_reverse} from Eq.~\ref{eq:explicit_reverse},
\begin{equation}
\mathbf{Z}_{t_k}^{\mathrm{exp}} - \mathbf{Z}_{t_k}^{\mathrm{imp}} = -h \cdot \Big[\, v_\theta(\mathbf{Z}_{t_{k+1}}, t_{k+1}) - v_\theta(\mathbf{Z}_{t_k}^{\mathrm{imp}}, t_k) \,\Big].
\label{eq:app_diff}
\end{equation}
Taylor-expanding the first term around $(\mathbf{Z}_{t_k}^{\mathrm{imp}}, t_k)$ and substituting the implicit relation $\mathbf{Z}_{t_{k+1}} - \mathbf{Z}_{t_k}^{\mathrm{imp}} = h\, v_\theta(\mathbf{Z}_{t_k}^{\mathrm{imp}}, t_k)$ yields
\begin{equation}
v_\theta(\mathbf{Z}_{t_{k+1}}, t_{k+1}) - v_\theta(\mathbf{Z}_{t_k}^{\mathrm{imp}}, t_k) = h \cdot \frac{D v_\theta}{D t}(\mathbf{Z}_{t_k}^{\mathrm{imp}}, t_k) + \mathcal{O}(h^2),
\end{equation}
where $\dfrac{D v_\theta}{D t} = \partial_t v_\theta + (\nabla_{\mathbf{z}} v_\theta)\, v_\theta$ is the material derivative of $v_\theta$ along its own flow. Plugging this into Eq.~\ref{eq:app_diff} gives the local deviation
\begin{equation}
\mathbf{Z}_{t_k}^{\mathrm{exp}} - \mathbf{Z}_{t_k}^{\mathrm{imp}} = -h^2 \cdot \frac{D v_\theta}{D t}(\mathbf{Z}_{t_k}^{\mathrm{imp}}, t_k) + \mathcal{O}(h^3).
\label{eq:app_local}
\end{equation}

\paragraph{Global error.}
Let $e_k = \mathbf{Z}_{t_k}^{\mathrm{exp}} - \mathbf{Z}_{t_k}^{\mathrm{imp}}$ denote the accumulated error, with $e_N = 0$. Applying Eq.~\ref{eq:app_local} together with the Lipschitz property of $v_\theta$ to propagate the error from step $k+1$ to step $k$ gives the recursion
\begin{equation}
\| e_k \| \leq (1 + h L)\, \| e_{k+1} \| + h^2 M + \mathcal{O}(h^3),
\end{equation}
where $M = \sup_{t \in [0,1]} \big\| \frac{D v_\theta}{D t}(\mathbf{Z}_t^{\mathrm{imp}}, t) \big\|$. Iterating this recursion from $k = N$ down to $k = 0$ and using $(1 + h L)^N \leq e^{L}$ with $N = 1/h$:
\begin{equation}
\| e_0 \| \leq h^2 M \sum_{j=0}^{N-1} (1 + h L)^j = h^2 M \cdot \frac{(1 + h L)^N - 1}{h L} \leq M\, h \cdot \frac{e^L - 1}{L} + \mathcal{O}(h^2).
\label{eq:app_global}
\end{equation}

\paragraph{Interpretation.}
Eq.~\ref{eq:app_global} shows that the accumulated deviation between explicit and exact implicit reverse sampling is linear in the step size $h$ and vanishes as $h \to 0$. The constant $M$ is small when the velocity field is nearly constant along its own flow which is the regime targeted by rectified-flow training and $L$ is bounded for any Lipschitz-constrained transformer backbone.

\section{Benchmarks and Datasets}
\label{app:datasets}

We evaluate physical understanding using three complementary datasets that probe different aspects of intuitive physics: IntPhys~\citep{DBLP:journals/corr/abs-1803-07616}, InfLevel~\citep{WeihsEtAl2022InfLevel}, and the controlled physics dataset of~\citet{kang2025farvideogenerationworld}.

\begin{figure}[t]
    \centering
    \includegraphics[width=\linewidth]{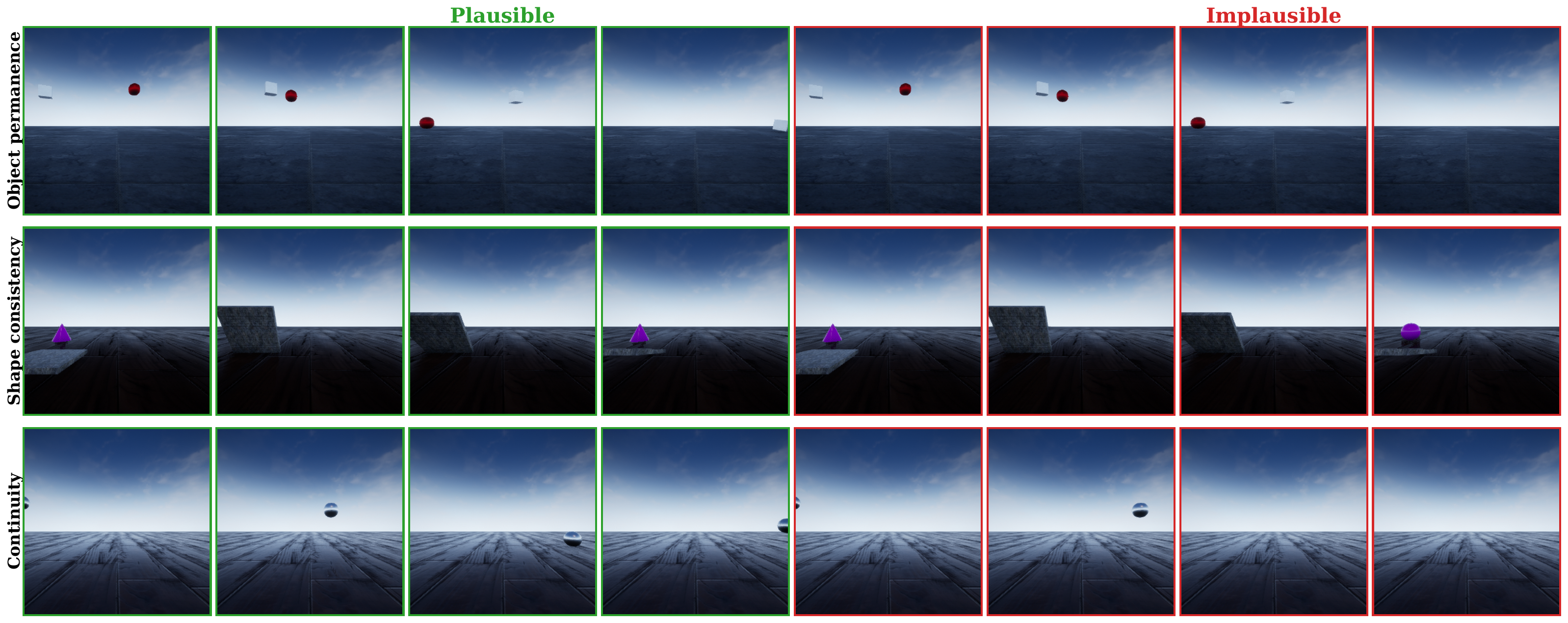}
    \caption{\textbf{IntPhys dataset examples.} Each pair shows a plausible (left) and an implausible (right) video under identical conditions, one pair per block: object permanence, shape constancy and spatio-temporal continuity.}
    \label{fig:inphys_examples}
\end{figure}

\paragraph{IntPhys.}
IntPhys (see~\Cref{fig:inphys_examples}) is designed around the violation-of-expectation paradigm from cognitive science, where models are asked to distinguish physically plausible from implausible events. The benchmark is organised in three blocks: object permanence (O1), shape constancy (O2) and spatio-temporal continuity (O3). Implausible videos contain violations such as an object vanishing behind an occluder, changing shape while out of view, or reappearing at a position inconsistent with its motion. Importantly, many of these violations require reasoning about hidden states (e.g., objects behind occlusion), making the task fundamentally different from simple motion pattern recognition.

\textbf{}

\begin{figure}[t]
    \centering
    \includegraphics[width=\linewidth]{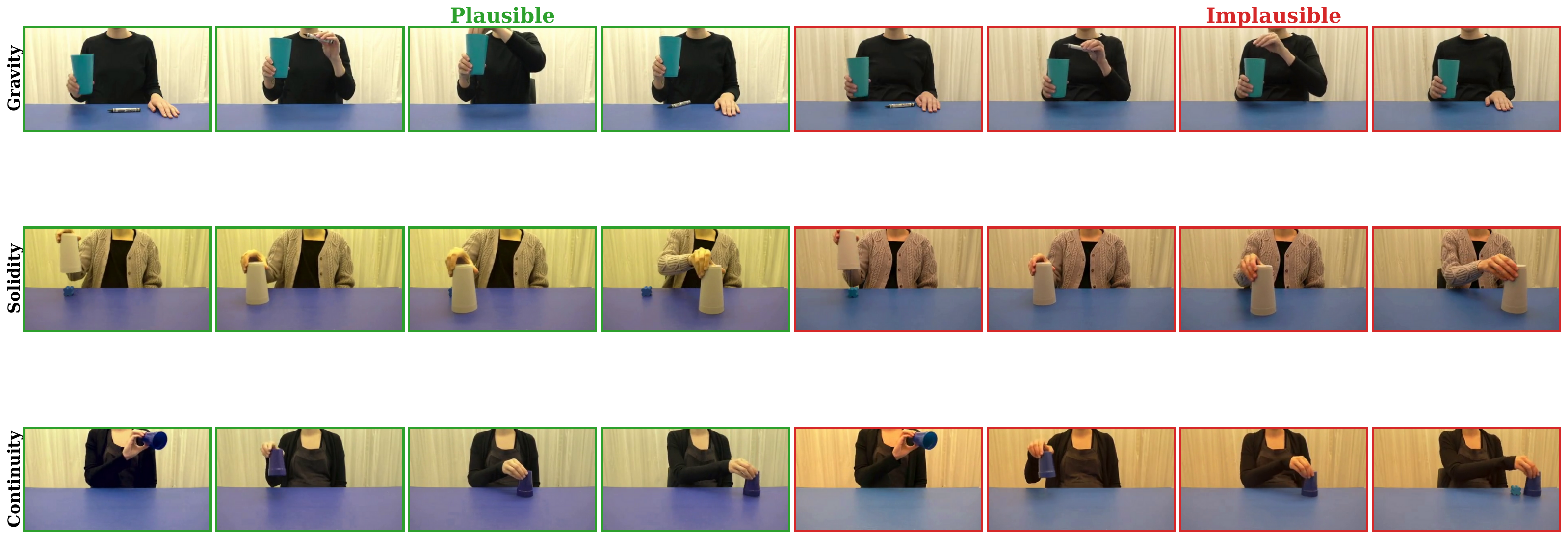}
    \caption{\textbf{InfLevel Dataset Examples.} Each pair shows a plausible (left) and implausible (right) video under identical conditions. Violations include gravity, solidity, and continuity, requiring reasoning beyond local motion cues.}
    \label{fig:inflevel_examples}
\end{figure}

\paragraph{InfLevel.}
InfLevel (see~\Cref{fig:inflevel_examples}) extends this paradigm by introducing more complex, compositional scenes with multiple interacting objects and longer temporal dependencies. In addition to basic physical constraints, InfLevel requires models to track object identities and interactions over time, often under partial observability. This increases the difficulty relative to IntPhys by requiring consistent reasoning across longer horizons and more cluttered environments. Although some scenarios involve gravitational motion, correctly modeling gravity is only one component of physical understanding. In these benchmarks, many violations are not detectable from local motion cues alone. For example, an object may move in a physically consistent way under gravity, yet violate object permanence by disappearing behind an occluder or reappearing inconsistently. Similarly, violations of solidity (objects intersecting) or support (objects floating without contact) require reasoning about spatial relationships and interactions rather than dynamics alone. 

\textbf{}

As a result, solving these benchmarks requires integrating multiple aspects of intuitive physics: (i) \emph{dynamics} (e.g., gravity and motion), (ii) \emph{object permanence} (tracking entities through occlusion), and (iii) \emph{interaction constraints} (e.g., collision, support, and non-penetration). This makes the task significantly more challenging than predicting trajectories in isolation, as the model must maintain a coherent internal representation of the scene over time.

\paragraph{Controlled physics dataset.}
To move beyond binary plausibility, we use the dataset of~\citet{kang2025farvideogenerationworld}, which is generated by a deterministic 2D physics simulator. Each video is associated with known physical parameters, including initial position, velocity, mass, and trajectory type. This allows us to evaluate whether internal representations encode quantitative physical variables, rather than simply distinguishing plausible from implausible outcomes.

\section{Internal structure}
\label{app:tsne}

\begin{figure}[ht]
    \centering
    \includegraphics[width=\linewidth]{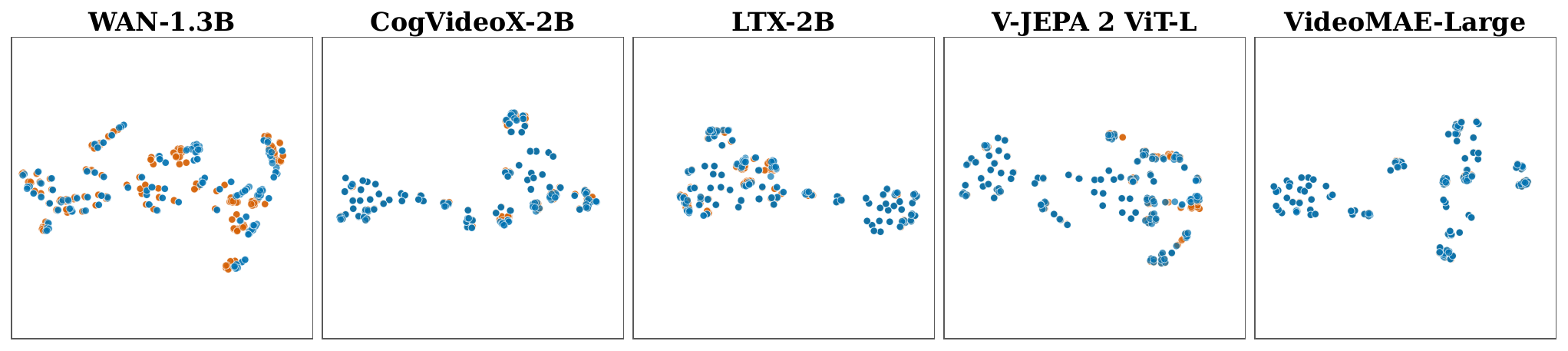}
    \caption{\textbf{t-SNE projections of internal activations on IntPhys.} Best-performing block at $t = 0.5$ for the three diffusion models, and best-performing block for V-JEPA 2 ViT-L and VideoMAE-Large. Plausible videos are shown in \textcolor{blue}{blue}, implausible in \textcolor{orange}{orange}.}
    \label{fig:tsne_internal}
\end{figure}

To complement the quantitative probing results in the main text, we visualise the internal representations of every model on IntPhys using t-SNE projections of the activations. Plausible videos are shown in blue and implausible videos in orange, and the same set of videos is projected through every model. The goal is to assess whether the physical plausibility distinction is also visible in an unsupervised geometric sense, without the help of a trained probe.
\begin{wrapfigure}{r}{0.45\linewidth}
    \vspace{-10pt}
    \centering
    \includegraphics[width=\linewidth]{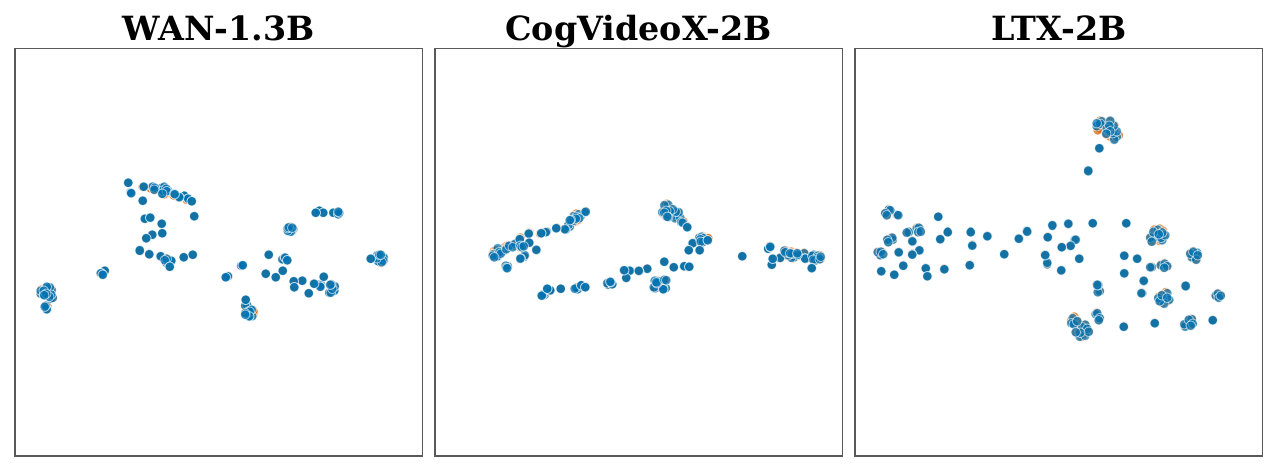}
    \vspace{-15pt}
    \caption{\textbf{t-SNE projections of VAE latents on IntPhys.} VAE latents $\mathbf{Z}_1$ before any flow computation. Plausible and implausible videos are intermixed, consistent with chance-level probe accuracy on this representation.}
    \label{fig:tsne_vae}
    \vspace{-10pt}
\end{wrapfigure}

\Cref{fig:tsne_internal} shows the projections at the best-performing block of each model. Some local separation between plausible and implausible clusters is visible, particularly for WAN-1.3B, but no model produces a clean unsupervised partition of the two classes. The structure is qualitatively consistent across diffusion models and dedicated representation encoders.
For comparison, \Cref{fig:tsne_vae} shows the same projection applied to the clean VAE latents $\mathbf{Z}_1$. The two colours are intermixed everywhere in the projection, consistent with the chance-level probe accuracy on VAE latents reported in \Cref{sec:plausibility}. The structure recovered by the probe at intermediate blocks is therefore not inherited from a structured input representation but is constructed by the diffusion model itself.

\section{Additional results on IntPhys 2}
\label{app:intphys2}

IntPhys~2~\citep{bordes2506intphys} extends the violation-of-expectation paradigm of IntPhys to photorealistic scenes with moving cameras and richer occlusions, and tests four conditions: permanence, immutability, solidity, and spatio-temporal continuity. It is considerably harder than its predecessor, and its authors report that most current video models perform close to chance. We apply the probing protocol of \Cref{sec:plausibility} without modification, and additionally include the larger V-JEPA~2 ViT-g and VideoMAE-Giant encoders as baselines.

\begin{figure}[h]
    \centering
    \includegraphics[width=\linewidth]{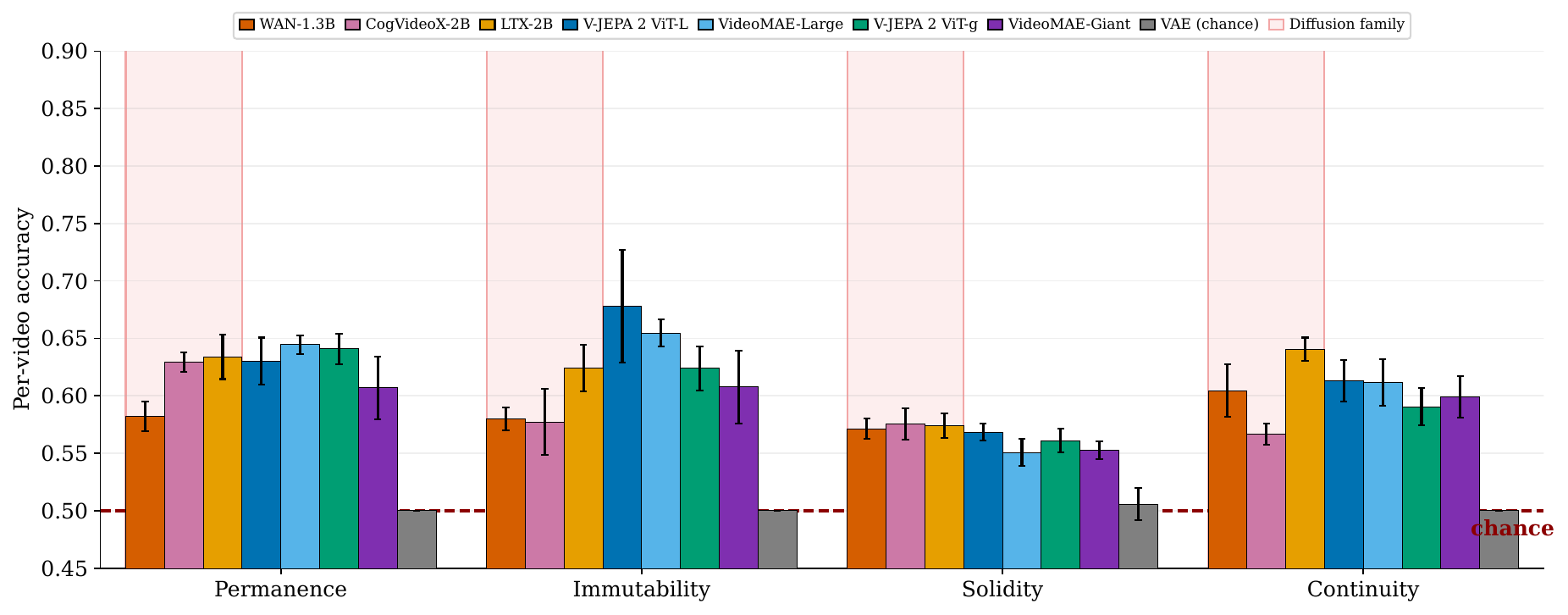}
    \caption{\textbf{Per-video probe accuracy on IntPhys~2.} Accuracy per condition for the three diffusion models, four representation encoders, and the VAE latents. Error bars show the standard error of the mean across 5 seeds.}
    \label{fig:intphys2}
\end{figure}

\Cref{fig:intphys2} reports per-video accuracy for each condition. All models fall well below their IntPhys and InfLevel accuracy, with every model between roughly $55\%$ and $68\%$. Within this range the diffusion models are on par with the representation encoders: they are marginally ahead on solidity, LTX-2B is the strongest model on continuity, and the encoders lead on immutability. As in the main experiments, the VAE latents remain at chance on all four conditions, so the signal the diffusion models carry is again constructed inside the transformer rather than inherited from the input. The ordering among diffusion models also differs from IntPhys and InfLevel, with LTX-2B rather than WAN-1.3B performing best.

\section{Structure of Learned Probe Weights}
\label{app:probe_weights}

In addition to reporting probe accuracy, we analyze the weights learned by the
linear probes to understand whether the physical plausibility signal has a
structured organization across denoising time and transformer depth. For each
transformer block $b$ and inversion noise level $\sigma$, the binary probe has
two class weight vectors, $w^{(0)}_{b,\sigma}$ for the implausible class and
$w^{(1)}_{b,\sigma}$ for the plausible class. We define the discriminative
probe direction as
\begin{equation}
    \Delta w_{b,\sigma} = w^{(1)}_{b,\sigma} - w^{(0)}_{b,\sigma}.
\end{equation}
The magnitude of this vector indicates how strongly the linear probe separates
plausible from implausible examples using the representation at that block and
noise level. We summarize this magnitude by averaging over feature dimensions,
$\frac{1}{d}\sum_i |\Delta w_{b,\sigma,i}|$.

\begin{figure}[t]
    \centering
    \begin{subfigure}{0.5\linewidth}
        \centering
        \includegraphics[width=\linewidth]{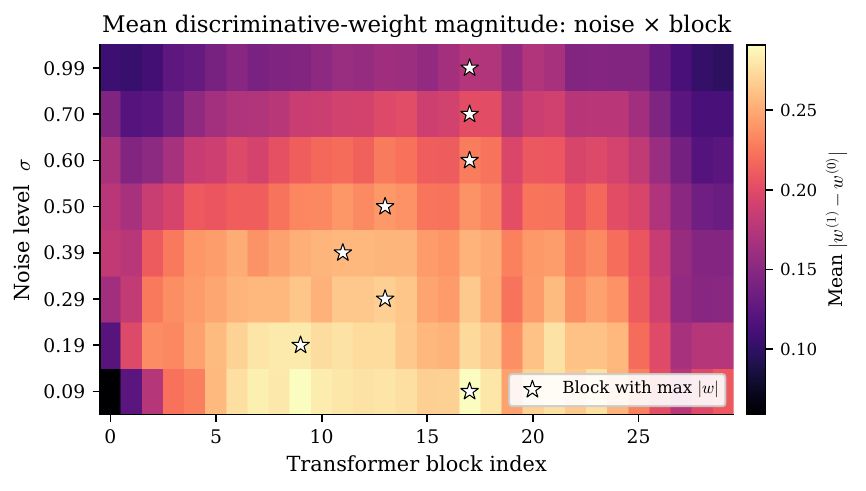}
        \caption{Noise $\times$ block map}
    \end{subfigure}
    \hfill
    \begin{subfigure}{0.48\linewidth}
        \centering
        \includegraphics[width=\linewidth]{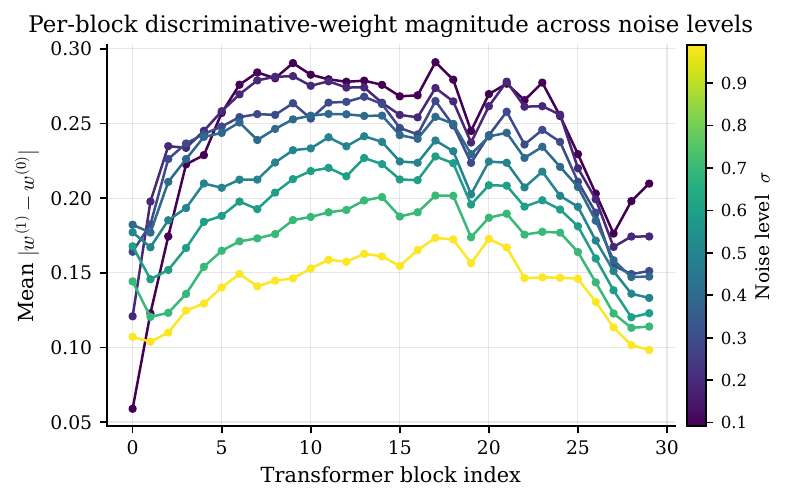}
        \caption{Block profiles}
    \end{subfigure}
    \caption{\textbf{Structure of learned probe weights.}
    For WAN on IntPhys, we plot the mean magnitude of the discriminative probe
    direction $\Delta w_{b,\sigma}=w^{(1)}_{b,\sigma}-w^{(0)}_{b,\sigma}$ across
    inversion noise level and transformer depth. The learned probe directions
    are strongest in a broad intermediate-depth region and vary smoothly across
    the denoising trajectory, matching the structure observed in the probe
    accuracy maps.}
    \label{fig:app_probe_weight_structure}
\end{figure}

These weight analyses shown in~\Cref{fig:app_probe_weight_structure} are complementary to the accuracy results in the main
text. Probe accuracy measures whether physical plausibility is linearly
decodable from a representation, whereas the weight magnitude measures the
strength and organization of the learned separating direction itself. The
consistent intermediate-depth structure in both views suggests that the
physical signal is not an artifact of a single probe or noise level, but is
organized systematically across the denoising computation.

\section{Reproducibility Details}
\label{app:reproducibility-details}

\paragraph{Datasets and splits.}
We evaluate on the validation/development splits of the physical-reasoning datasets used in the paper. For the probe experiments, we split the extracted scene-level features into train and validation subsets with a fixed random seed of 42. The validation fraction is 40\%, and all reported probe accuracies are computed on this held-out validation subset. 

\paragraph{Inference and feature extraction.}
For each scene, we run deterministic reverse sampling with the pretrained model weights and save transformer block activations at the requested denoising steps. Unless otherwise stated, videos are resized to $256 \times 256$. WAN and CogVideoX are run with 81 frames, while LTX-Video is run with 97 frames to satisfy the model's frame-count constraint. The default 100-step setting uses classifier-free guidance scale 1.0 and captures the requested inversion step, step 50 ($t=0.5$) for the final-step probe analysis. WAN experiments use 30 transformer blocks with hidden dimension 1536, CogVideoX uses 30 blocks with hidden dimension 1920, and LTX-Video uses 28 blocks with hidden dimension 2048. 

\paragraph{Linear probes.}
For each model, dataset, and denoising step, we train one linear probe per transformer block. Each saved block-output tensor is mean-pooled over tokens before the probe, giving one feature vector per block. Each probe is a linear classifier from the model hidden dimension to the binary plausibility label. The loss is the mean cross-entropy across all block probes. We train for 50 epochs with Adam, learning rate $10^{-3}$, batch size 4. The validation metrics are computed on the fixed 40\% validation split described above. We report per-block validation accuracy, per-task validation accuracy by grouping held-out scenes according to their task family.

\paragraph{Noise intervention experiment.}
For the probe-surprise intervention analysis, we start from plausible scenes with saved recovered noise latents and regenerate a baseline video. We then repeat generation while intervening on one transformer block at a time. Gaussian noise is added to the selected block activations across the denoising trajectory with intervention strength $\alpha=0.5$. We sweep all transformer blocks for the selected model and score each generated video by re-inverting it, extracting activations at the requested probe step, and applying the corresponding trained probe checkpoint. For each intervened block, we report the change in average probe surprise relative to the non-intervened baseline. The main noise-intervention runs use the step-50 probe checkpoint and evaluate all available plausible scenes in the selected dataset split.

\paragraph{Hardware and Compute Times. }
All inference, probing, and intervention experiments were executed on cluster nodes using a single NVIDIA GH200 GPU per job (96GB memory).

\end{document}